\documentclass{ws-procs975x65}
\newcommand{\ba}{\begin{eqnarray}}
\newcommand{\ea}{\end{eqnarray}}

\begin{document}

\title{Partial Dynamical Symmetry as an\\ 
Intermediate Symmetry Structure}

\author{ A. Leviatan}

\address{Racah Institute of Physics, The Hebrew University,
Jerusalem 91904, Israel\\
E-mail: ami@vms.huji.ac.il}

\maketitle

\abstracts{
We introduce the notion of a partial dynamical symmetry for which a 
prescribed symmetry is neither exact nor completely broken. 
We survey the different types of partial dynamical symmetries and present 
empirical examples in each category.}

\section{Introduction}

Symmetries play an important role in dynamical systems. They provide 
quantum numbers for the classification of states, determine selection rules 
and facilitate the calculation of matrix elements.  
An exact symmetry occurs when the Hamiltonian of the system commutes 
with all the generators ($g_i$) of the symmetry-group,  
$[\, H \, , \, g_i\,] = 0$. 
In this case, all states have good symmetry and are labeled by the 
irreducible representations (irreps) of the group. 
The Hamiltonian admits a block structure so that 
inequivalent irreps do not mix and all eigenstates 
in the same irrep are degenerate. In a dynamical symmetry the block 
structure of the Hamiltonian is retained, the states preserve 
the good symmetry but in general are no longer degenerate 
(splitting but no mixing). When the symmetry is completely broken 
$[\, H \, , \, g_i\,] \neq 0$, and none 
of the states have good symmetry. In-between these limiting cases there 
may exist intermediate symmetry structures, called partial (dynamical) 
symmetries for which the symmetry is neither exact nor completely broken.

Models based on spectrum generating algebras, such as those 
developed \cite{ibm,vibron} by 
F. Iachello and his colleagues, form a convenient framework for 
discussing these different types of symmetries. 
In such models the Hamiltonian is written in terms of the generators of 
a Lie algebra, called the spectrum generating algebra. 
A dynamical symmetry occurs if the Hamiltonian
can be written in terms of the Casimir operators ($\hat{C}_{G_i}$)
of a chain of nested algebras
\ba
\begin{array}{lllllll}
G_1 & \supset & G_2 & \supset & \ldots & \supset & G_n  \\
\left [\alpha_1\right ] && \left [\alpha_2 \right ]&& 
\ldots  && \left [\alpha_n \right ]\,
\end{array}
\label{ds}
\ea
in which case it has the following properties:
(i)~{\it solvability:}
all states are solvable and analytic expressions
are available for energies and other observables;
(ii)~{\it quantum numbers:}
all states are classified by quantum numbers 
$\alpha_1,\, \alpha_2,\, \ldots\, \alpha_n$, 
which are the labels of the irreps
of the algebras in the chain;
(iii)~{\it pre-determined structure:}
the structure of wave functions is completely dictated by symmetry
and is independent of the Hamiltonian's parameters $a_i$
\ba
H &=& a_1\,\hat C_{G_1} + a_2\,\hat C_{G_2} 
+ \, \ldots \; + \; a_n\,\hat C_{G_n} ~.
\ea
The merits of a dynamical symmetry are self-evident.
However, in most applications to realistic systems,
the predictions of an exact dynamical symmetry are rarely fulfilled
and one is compelled to break it. This is usually done 
by including in the Hamiltonian symmetry-breaking terms associated with 
different sub-algebra chains of the parent spectrum generating algebra 
($G_1$). 
In general, under such circumstances, solvability is lost,
there are no remaining non-trivial conserved quantum numbers and all
eigenstates are expected to be mixed.
A partial dynamical symmetry (PDS) corresponds to
a particular symmetry breaking for which some (but not all) of the above
mentioned virtues of a dynamical symmetry are retained. 
It is then possible to identify the following types of 
partial dynamical symmetries: 
\begin{itemize}
\item {\em type I:} 
$\qquad\;\;\;$ {\bf part} of the states have {\bf all} the 
dynamical symmetry
\item{\em type II:}
$\qquad\;\,$ {\bf all} the states have {\bf part} of the 
dynamical symmetry
\item{\em type III:}
$\qquad$ {\bf part} of the states have {\bf part} of the dynamical 
symmetry
\end{itemize}
In what follows we explain each type of partial symmetry and show an 
empirical example of it. For that purpose we use the 
interacting boson model \cite{ibm} (IBM) 
based on a $U(6)$ spectrum generating algebra. The model 
describes low-lying quadrupole collective states in even-even nuclei
in terms of a system of $N$ monopole ($s$) and quadrupole ($d$) bosons
representing valence nucleon pairs.

\section{SU(3) PDS (type I)}

\begin{figure}[t]
\begin{center}
\epsfxsize=25pc
\epsfbox{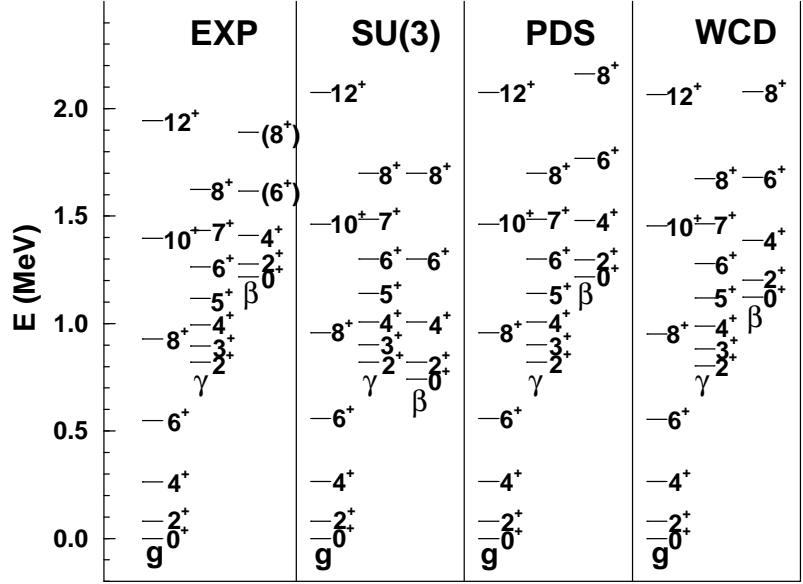}
\caption{Spectra of $^{168}$Er ($N=16$). Experimental energies
(EXP) are compared~\protect\cite{lev96} 
with IBM calculations in an exact $SU(3)$ dynamical
symmetry [$SU(3)$], in a $SU(3)$ PDS 
with a Hamiltonian $H+\lambda_1 L\cdot L$, Eq.~(\ref{hsu3}), and parameters 
$t_0=2t_2=4,\,\lambda_1=13$ keV (PDS), and in a broken $SU(3)$ 
symmetry (WCD)~{\protect\cite{WCD}} 
where an $O(6)$ term is added to an $SU(3)$ Hamiltonian.}
\end{center}
\end{figure}
\begin{figure}[t]
\begin{center}
\vspace{-2cm}
\epsfig{file=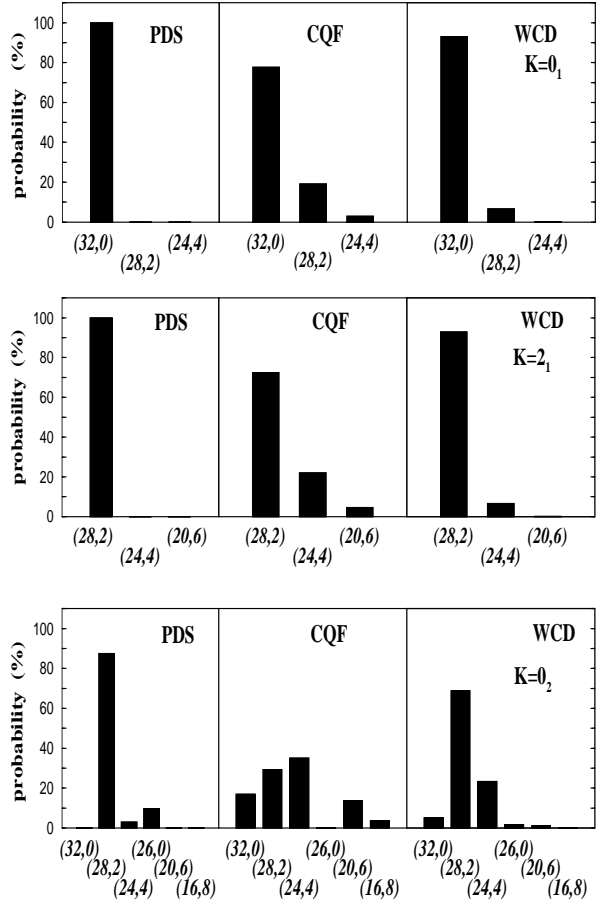,height=17cm,width=\linewidth,angle=0}
\vspace{-3.8cm}
\caption{
$SU(3)$ decomposition of wave functions of the ground ($K=0_1$), 
$\gamma$ ($K=2_1$),
and $K=0_2$ bands of $^{168}$Er ($N=16$) in the reported SU(3) PDS 
calculation~{\protect\cite{levsin99}}, 
and in broken-SU(3) calculations: 
CQF~{\protect\cite{CQF}} with a non-$SU(3)$ quadrupole operator in 
the Hamiltonian, and WCD~{\protect\cite{WCD}}.
\label{su3decomp}}
\end{center}
\end{figure}
\begin{table}[t]
\caption{
$B(E2)$ branching ratios from states in the $\gamma$ band in
$^{168}$Er. The column EXP is the experimental ratios, 
WCD is the broken $SU(3)$ calculation {\protect\cite{WCD}} and 
PDS is the reported $SU(3)$ partial dynamical symmetry 
calculation {\protect\cite{lev96}}.}
\begin{center}
\footnotesize
\begin{tabular}{lcccc|rlcccc}
\hline
$L^{\pi}_{i}$ & $L^{\pi}_{f}$ &  EXP &  PDS &  WCD &    &
$L^{\pi}_{i}$ & $L^{\pi}_{f}$ &  EXP &  PDS &  WCD \\
\hline
$2^{+}_{\gamma}$ & $0^{+}_{g}$      & $54.0$   &  $64.27$ &  $66.0$  &    &
$6^{+}_{\gamma}$ & $4^{+}_{g}$      &   $0.44$ &   $0.89$ &   $0.97$ \\
                 & $2^{+}_{g}$      & $100.0$  & $100.0$  & $100.0$  &    &
                 & $6^{+}_{g}$      &   $3.8$  &   $4.38$ &   $4.3$  \\
                 & $4^{+}_{g}$      &   $6.8$  &   $6.26$ &   $6.0$  &    &
                 & $8^{+}_{g}$      &   $1.4$  &   $0.79$ &   $0.73$ \\
$3^{+}_{\gamma}$ & $2^{+}_{g}$      &   $2.6$  &   $2.70$ &   $2.7$  &    &
                 & $4^{+}_{\gamma}$ & $100.0$  & $100.0$  & $100.0$  \\
                 & $4^{+}_{g}$      &   $1.7$  &   $1.33$ &   $1.3$  &    &
                 & $5^{+}_{\gamma}$ &  $69.0$  &  $58.61$ &  $59.0$  \\
                 & $2^{+}_{\gamma}$ & $100.0$  & $100.0$  & $100.0$  &    &
$7^{+}_{\gamma}$ & $6^{+}_{g}$      &   $0.74$ &   $2.62$ &   $2.7$  \\
$4^{+}_{\gamma}$ & $2^{+}_{g}$      &   $1.6$  &   $2.39$ &   $2.5$  &    &
                 & $5^{+}_{\gamma}$ & $100.0$  & $100.0$  & $100.0$  \\
                 & $4^{+}_{g}$      &   $8.1$  &   $8.52$ &   $8.3$  &    &
                 & $6^{+}_{\gamma}$ &  $59.0$  &  $39.22$ &  $39.0$  \\
                 & $6^{+}_{g}$      &   $1.1$  &   $1.07$ &   $1.0$  &    &
$8^{+}_{\gamma}$ & $6^{+}_{g}$      &   $1.8$  &   $0.59$ &   $0.67$ \\
                 & $2^{+}_{\gamma}$ & $100.0$  & $100.0$  & $100.0$  &    &
                 & $8^{+}_{g}$      &   $5.1$  &   $3.57$ &   $3.5$  \\
$5^{+}_{\gamma}$ & $4^{+}_{g}$      &   $2.91$ &   $4.15$ &   $4.3$  &    &
                 & $6^{+}_{\gamma}$ & $100.0$  & $100.0$  & $100.0$  \\
                 & $6^{+}_{g}$      &   $3.6$  &   $3.31$ &   $3.1$  &    &
                 & $7^{+}_{\gamma}$ & $135.0$  &  $28.64$ &  $29.0$  \\
                 & $3^{+}_{\gamma}$ & $100.0$  & $100.0$  & $100.0$  &    &
                 &                  &          &          &          \\
                 & $4^{+}_{\gamma}$ & $122.0$  &  $98.22$ &  $98.5$  \\
                 &                  &          &          & & & & & & &  \\
\hline
\end{tabular}
\end{center}
\end{table}
Partial dynamical symmetry of the first type corresponds to a situation for 
which {\it part} of the states preserve {\it all} the dynamical symmetry.
In this case the properties of solvability, good quantum numbers,
and symmetry-dictated structure are fulfilled exactly, but by only a 
subset of states. As an example we consider 
the IBM chain
\ba
\begin{array}{ccccc}
U(6) &\supset& SU(3) &\supset& O(3) \\
 \left [N\right ] && (\lambda,\mu)& K & L 
\end{array}
\label{dssu3}
\ea
applicable to axially deformed nuclei. A rotational-invariant 
IBM Hamiltonian with partial $SU(3)$ symmetry has the form \cite{lev96}
\ba
H \;=\; t_{0}\,\Gamma^{\dagger}_{0}\Gamma_{0} 
+ t_{2}\,\Gamma^{\dagger}_{2}\cdot\tilde \Gamma_{2}
\label{hsu3}
\ea
It consists of boson-pairs
\ba
\Gamma^{\dagger}_{0} =  d^{\dagger}\cdot d^{\dagger} - 2\,(s^{\dagger})^2
\; , \;
\Gamma^{\dagger}_{2,\mu} = 2\,s^{\dagger}d^{\dagger}_{\mu}
+ \sqrt{7}(d^{\dagger}d^{\dagger})^{(2)}_{\mu}
\label{su3pair}
\ea
which are $SU(3)$ tensors with $(\lambda,\mu) = (0,2)$ and $L=0,2$. 
For $t_0=t_2$ the above Hamiltonian is related to 
the Casimir operator of $SU(3)$, hence 
has an exact $SU(3)$ symmetry. For $t_0\neq t_2$, $H$ involves a mixture of 
$SU(3)$ tensors with $(\lambda,\mu) = (0,0)\oplus (2,2)$ and although 
it is not an $SU(3)$ scalar, it has a subset of solvable states with 
good $SU(3)$ symmetry. This arises from the fact that the boson pairs 
of Eq.~(\ref{su3pair}) satisfy $\Gamma_{L,\mu}\vert c;\,N \rangle = 0$, where 
\ba
\vert c;\,N \rangle &=& (N!)^{-1/2}(b^{\dagger}_{c})^{N}\vert 0 \rangle
\; , \;
b^{\dagger}_c = (\sqrt{2}\,d^{\dagger}_{0} +s^{\dagger} )/\sqrt{3}
\label{condsu3}
\ea
is the lowest weight state in the $SU(3)$ irrep 
$(\lambda,\mu) = (2N,0)$. In addition, 
$[\, \Gamma_{L,\mu}\, , \, \Gamma^{\dagger}_{2,2}\,]
\vert c; \, N \rangle \propto \delta_{L,2}\,\delta_{\mu,2}\,
\vert c;\,N \rangle$ and 
$[\,[\, \Gamma_{L,\mu}\, , \, \Gamma^{\dagger}_{2,2}\,]\, , \,
\Gamma^{\dagger}_{2,2}\,] \propto 
\delta_{L,2}\,\delta_{\mu,2}\, \Gamma^{\dagger}_{2,2}$, 
from which it follows that the sequence of states
$\vert k\rangle = (\Gamma^{\dagger}_{2,2})^{k}\vert c;\,N-2k \rangle$
are eigenstates of $H$ with good $SU(3)$ symmetry 
$(\lambda,\mu)=(2N-4k,2k)$. The states $\vert k\rangle$ are deformed 
and serve as intrinsic states representing the ground band ($k=0$) and 
$\gamma^k$ bands with angular momentum projection $K=2k$ along the symmetry 
axis. Since the Hamiltonian $H$ of Eq.~(\ref{hsu3}) is an $O(3)$ scalar, 
the rotational states projected from these intrinsic states are also 
solvable eigenstates of $H$ with good $SU(3)$ symmetry. States in other bands 
are mixed. 
Adding to $H$ $O(3)$ rotation terms produces an $L(L+1)$ 
splitting and lead to a $SU(3)$ PDS of type I. 
The corresponding spectrum is shown in Fig.~1 
in comparison with $^{168}$Er, and 
the $SU(3)$ decomposition of the lowest bands is given in Fig.~2. 
The ground ($K=0_1$) and $\gamma$ ($K=2_1$) 
bands are solvable with good $SU(3)$ symmetry $(\lambda,\mu)=(2N,0)$ 
and $(2N-4,2)$ respectively. 
Unlike the case of an exact dynamical symmetry, the  
first $K=0_2$ band is no longer degenerate with the $\gamma$-band, 
in agreement with the empirical situation in most deformed nuclei. 
Futhermore, the $K=0_2$ band involves a mixture of $SU(3)$ irreps 
$(2N-4,2)\oplus (2N-8,4)\oplus (2N-6,0)$ or equivalently a mixture of 
a single-phonon $(87.5\%\;\beta)$ and double-phonon 
$(12.4\%\;\gamma^2_{K=0}$ and $0.1\%\; \beta^2)$ 
components \cite{levsin99}.

Electromagnetic transitions provide a sensitive test for the 
structure of states. As shown in Table 1, 
the $SU(3)$ PDS E2 rates for transitions originating from the 
$\gamma$ band are found to be in excellent agreement with experiment. 
The calculated values are obtained by using 
the general IBM E2 operator 
$T^{(2)} = \alpha\,Q^{(2)} + \theta\,(d^{\dagger}s 
+ s^{\dagger}\tilde{d})$. $Q^{(2)}= d^{\dagger}s 
+ s^{\dagger}\tilde d -(\sqrt{7}/2)(d^{\dagger}\tilde d)^{(2)}$ 
is an $SU(3)$ generator, hence cannot connect 
the ground and $\gamma$ bands which have 
different $SU(3)$ character.
This property combined with the fact that the corresponding wave 
functions of these solvable bands are determined solely by symmetry, imply 
that the $B(E2)$ ratios for $\gamma\to g$ transitions quoted in Table 1 
do not depend on parameters of the $E2$ operator nor of the Hamiltonian 
and therefore are parameter-free predictions of $SU(3)$ PDS. 
The agreement between these predictions and the data confirms the 
relevance of $SU(3)$ PDS to the spectroscopy of $^{168}$Er.

\section{O(6) PDS (type I)}

It is possible to apply a similar procedure to construct a Hamiltonian 
with a partial symmetry for the chain
\ba
\begin{array}{ccccccc}
U(6) &\supset& O(6) &\supset& O(5) &\supset& O(3) \\
 \left [N\right ] && \langle0,\sigma,0\rangle && (\tau,0) && L
\end{array}
\label{dsO6}
\ea
The $O(6)$ intrinsic state for the ground band 
\ba
\vert c;\,N \rangle &=& (N!)^{-1/2}(b^{\dagger}_{c})^{N}\vert 0 \rangle
\; , \;
b^{\dagger}_c = (\,d^{\dagger}_{0} + s^{\dagger} )/\sqrt{2}
\label{condeno6}
\ea
has $\sigma=N$ and the boson pair which annihilates it, 
$P_{0}\vert c;\,N \rangle = 0$, has the form
\ba
P^{\dagger}_{0} &=&  d^{\dagger}\cdot d^{\dagger} - \,(s^{\dagger})^2 ~.
\label{O6p0}
\ea
The resulting Hamiltonian, 
$H_{O(6)} = A\,P^{\dagger}_{0}P_{0}$ is related to the 
Casimir operator of $O(6)$, hence has an exact $O(6)$ symmetry. 
Adding to it the $O(5)$ and $O(3)$ Casimir operators induces 
$\tau(\tau+3)$ and $L(L+1)$ splitting and lead to an 
$O(6)$ dynamical symmetry.  
The latter has been used \cite{ciz79} to describe the structure of 
the $\gamma$-unstable deformed nucleus $^{196}$Pt.
The agreement is excellent for properties of the ground band 
($\sigma=N$), yet the resulting fit for the observed anharmonicity of 
excited bands is quite poor. 
In the dynamical symmetry limit the lowest bands have $\sigma=N,N-2,N-4$ 
and the eigenvalues $A\,(N-\sigma)(N+\sigma+4)$ of $H_{O(6)}$ imply a 
fixed anharmonicity: $2 [\, 1 - {1\over N+1}\, ]$. 
For $^{196}$Pt with $N=6$ , the predicted 
anharmonicity is $1.71$ compared to the empirical value $1.30$. 
One is therefore motivated to search for a Hamiltonian which will improve 
the fit to the intrinsic spectrum without destroying the good $O(6)$ 
description for the ground band. 
This can be accomplished \cite{levramisa03} by the 
following Hamiltonian with an $O(6)$ PDS of type II 
\ba
H = r_{0}\,R^{\dagger}_{0}R_{0} 
+ r_{2}\,R^{\dagger}_{2}\cdot\tilde R_{2} ~.
\label{h3bod}
\ea
The boson-triplets 
\ba
R^{\dagger}_{0} =  s^{\dagger}P^{\dagger}_{0} \;\; , \;\; 
R^{\dagger}_{2,\mu} =  d^{\dagger}_{\mu}P^{\dagger}_{0}
\ea
are $O(6)$ tensors with $\sigma=1$. For $r_0=r_2$, 
the Hamiltonian $H$ is proportional to $H_{O(6)}$ hence has an exact 
$O(6)$ symmetry. 
For $r_0\neq r_2$ it involves a mixture of $O(6)$ tensors with 
$(\sigma=0)\oplus (\sigma=2)$. 
In general, although $H$ is not an $O(6)$ scalar, it satisfies by 
construction $H\vert c;\,N \rangle = 0$, and therefore has an exactly
solvable ground band with good $O(6)$ symmetry $\sigma=N$.
Since $H$ is an $O(5)$ scalar, states of good $O(5)$ symmetry $\tau$ and 
good angular momentum $L$ projected from $\vert c;\,N\rangle$ are also 
eigenstates of $H$ and form a ground band endowed with good $O(6)$ dynamical 
symmetry. In contrast, states in excited bands mix several $\sigma$ irreps. 
Clearly, the Hamiltonian (\ref{h3bod}) with added $O(5)$ and $O(3)$ 
rotational terms exhibits $O(6)$ PDS of type~I. 
Preliminary calculations \cite{levramisa03} indicate that 
such Hamiltonian preserves the good $O(6)$ description 
for the ground band and is able reproduce the empirical anharmonicity 
of excited bands in $^{196}$Pt.

It is also possible to consider a partial dynamical symmetry with respect 
to the third IBM chain: 
$U(6) \supset U(5) \supset O(5) \supset O(3)$ with quantum numbers 
$N,n_d,\tau,L$ respectively. A three-body Hamiltonian with a $U(5)$ 
PDS of type I was presented by Talmi \cite{talmi03}. 
A general algorithm how to construct Hamiltonians with partial 
dynamical symmetry of type I 
for any semi-simple group is available~\cite{alha92}. 

\section{O(5) PDS (type II)}

The second type of PDS corresponds to a situation for which {\it all} the
states preserve {\it part} of the dynamical symmetry.
In this case there are no analytic solutions,
yet selected quantum numbers (of the conserved symmetries) are retained.
This occurs, for example, when the Hamiltonian contains interaction
terms from two different chains with
a common symmetry subalgebra, as in the following IBM 
chains~\cite{levnov86}
\ba
\left.
\begin{array}{ccc}
U(6) & \supset & U(5) \\
U(6) & \supset & O(6) \\
\end{array}
\right\} \supset O(5) \supset O(3) ~.
\label{u5o6}
\ea
A realization of such an $O(5)$ PDS of type II, is given by the 
following Hamiltonian, typical for the $U(5)$ (spherical) to O(6) 
(deformed $\gamma$-unstable) transition region 
\ba
H = \epsilon\,\hat{n}_d 
+ A\, P^{\dagger}_{0} P_{0} ~.
\label{hu5o6}
\ea
Here $\hat{n}_d$ is the $d$-boson number operator which is a Casimir 
operator of $U(5)$ and the $A$-term is the $O(6)$ pairing term mentioned 
in Eq.~(\ref{O6p0}). In this case, 
all eigenstates of $H$ have good $O(5)$ symmetry but none of them 
have good $U(5)$ nor good $O(6)$ symmetries and hence only part 
of the dynamical symmetry of each chain in Eq.~(\ref{u5o6}) is observed. 
The $E(5)$ critical point of the second order shape-phase transition, 
considered recently by Iachello~\cite{iac00}, 
correspond to the Hamiltonian of Eq.~(\ref{hu5o6}) with
$\epsilon = (N-1)\, A$, and falls into the present PDS category. 

\section{O(6) PDS (type II)}

\begin{figure}[t]
\begin{center}
\epsfxsize=30pc
\epsfig{file=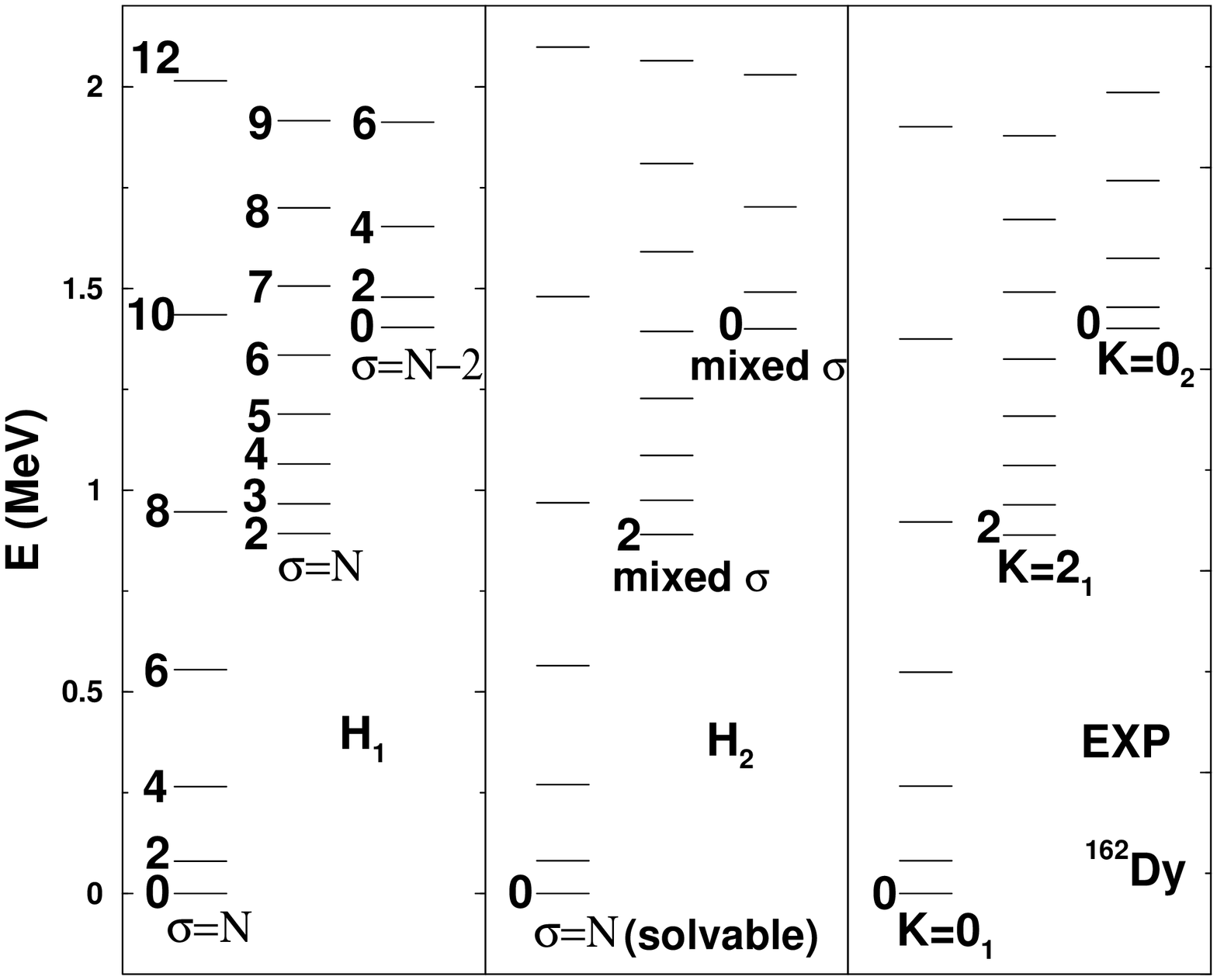,width=\linewidth,angle=270}
\caption{Experimental spectra
(EXP) of $^{162}$Dy 
compared with calculated spectra~\protect\cite{levisa02} of
$H_1+\lambda_1 L\cdot L$, Eq.~(\ref{h1}), and
$H_2+\lambda_2 L\cdot L$, Eq.~(\ref{h2}), with parameters (in keV)
$\kappa_0=8$, $\kappa_2=1.364$, $\lambda_1=8$ and
$h_0=28.5$, $h_2=6.3$, $\lambda_2=13.45$
and boson number $N=15$.
\label{so6energy}}
\end{center}
\end{figure}
An alternative situation where PDS of type II can occur is 
when the Hamiltonian preserves only some of the symmetries $G_i$ in the 
chain~(\ref{ds}), and only their irreps are unmixed.
Such a scenario was recently 
considered by Van Isacker~\cite{isa99} in relation to the $O(6)$ 
chain of Eq.~(\ref{dsO6}), using the following Hamiltonian
\ba
H_1 \;=\; \kappa_{0}P^{\dagger}_{0}P_{0}
+ \kappa_2 \Bigl (\Pi^{(2)}\times \Pi^{(2)}\Bigr )^{(2)}\cdot\Pi^{(2)} ~.
\label{h1}
\ea
The $\kappa_0$ term is the $O(6)$ pairing term mentioned in 
Eq.~(\ref{O6p0}).
The $\kappa_2$ term is constructed only from the $O(6)$ generator,
$\Pi^{(2)}=d^{\dagger}s+s^{\dagger}\tilde{d}$,
which is not a generator of $O(5)$.
Therefore, it cannot connect states in different $O(6)$ irreps
but can induce $O(5)$ mixing subject to $\Delta\tau=\pm 1,\pm 3$.
Consequently, $H_1$ preserves the $U(6)$, $O(6)$, and $O(3)$ symmetries 
(with good quantum numbers $N,\sigma,L$) but not the $O(5)$ symmetry 
(and hence leads to $\tau$ admixtures).
These are the necessary ingredients of an $O(6)$ PDS of type II associated 
with the chain in Eq.~(\ref{dsO6}).

In Fig.~3 we show the experimental spectrum of $^{162}$Dy
and compare with the calculated spectra of $H_1$ (\ref{h1}).
The spectra display rotational bands of an axially-deformed nucleus,
in particular, a ground band $(K=0_1)$
and excited $K=2_1$ and $K=0_2$ bands.
As shown in the upper portion of Fig.~4, 
all bands of $H_1$ are pure with respect to $O(6)$.
Specifically, the $K=0_1,2_1,2_3$ bands have $\sigma=N$
and the $K=0_2$ band has $\sigma=N-2$.
In this case the diagonal $\kappa_0$-term in Eq.~(\ref{h1}) simply
shifts each band as a whole in accord with its $\sigma$ assignment.
On the other hand, the $\kappa_2$-term in Eq.~(\ref{h1}) 
is an $O(5)$ tensor with $\tau=3$ and, therefore, 
all eigenstates of $H_1$ are mixed with respect to 
$O(5)$. This mixing is demonstrated in the upper portion of Fig.~5 
for the $L=0,2$ members of the ground band.
\noindent
\begin{figure}[t]
\begin{minipage}{0.48\linewidth}
\epsfig{file=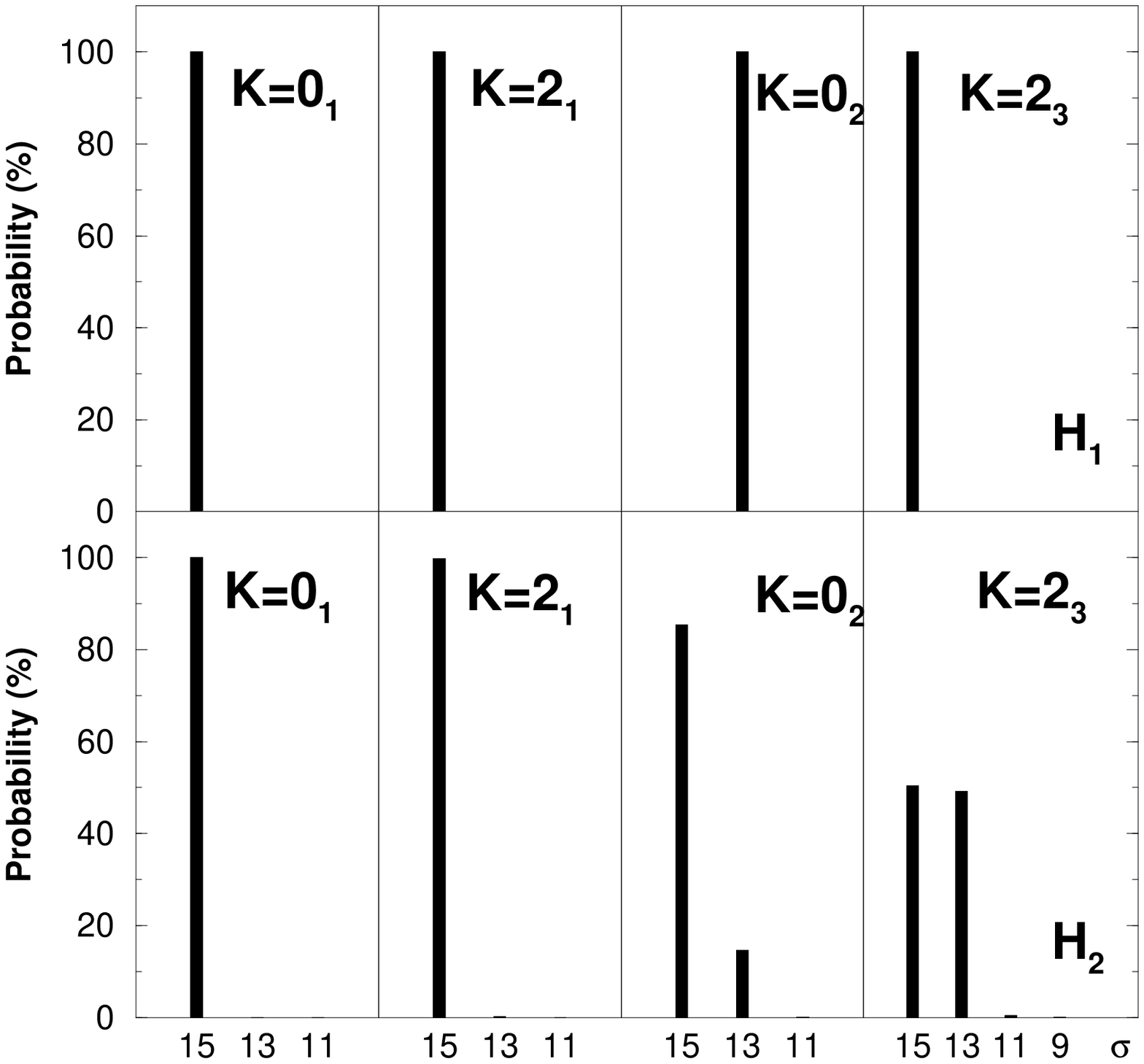,height=65mm,width=\linewidth,angle=270}
\vspace{-0.45cm}
\caption{~$O(6)$ decomposition~\protect\cite{levisa02} of wave 
functions of states in the bands 
$K=0_1,\,2_1,\,0_2,\,(L=K^{+})$, and $K=2_3,\,(L=3^{+})$, for 
$H_1$ (upper portion) and $H_2$ (lower portion).}
\end{minipage}
\hspace{\fill}
\begin{minipage}{0.48\linewidth}
\hspace{2cm}
\epsfig{file=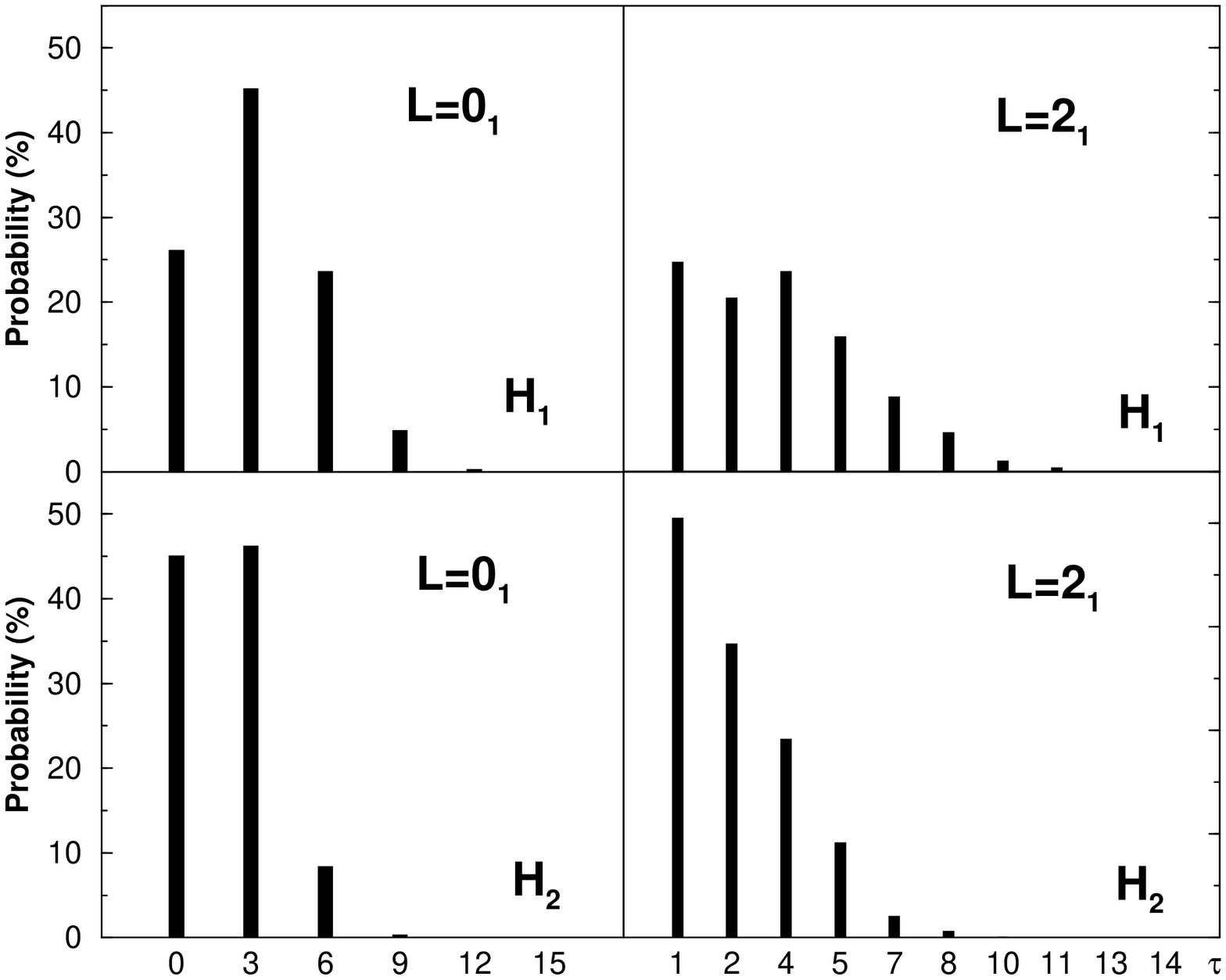,height=65mm,width=\linewidth,angle=270}
\caption{ 
$O(5)$ decomposition~\protect\cite{levisa02} of wave 
functions of the $L=0,\,2$ states
in the ground band ($K=0_1$) of $H_1$ (\ref{h1}) [upper portion] and
$H_2$ (\ref{h2}) [lower portion]. Both states have $\sigma=N$.}
\end{minipage}
\end{figure}
\section{O(6) PDS (type III)}

The third type of partial symmetries 
has a hybrid character, for which {\it part} of the states of the 
system under study preserve {\it part} of the dynamical symmetry.
Such a generalized partial symmetry associated with the 
$O(6)$ chain of Eq.~(\ref{dsO6}), can be realized by 
the Hamiltonian \cite{levisa02}
\ba
H_2 \;=\; h_{0}P^{\dagger}_{0}P_{0} + h_{2}P^{\dagger}_{2}
\cdot\tilde P_{2} ~.
\label{h2}
\ea
Here $P^{\dagger}_{0}$ is the $\sigma=0$ pair of Eq.~(\ref{O6p0}) 
and the second boson-pair
\ba
P^{\dagger}_{2,\mu} = \sqrt{2}\,s^{\dagger}d^{\dagger}_{\mu}
+ \sqrt{7}(d^{\dagger}d^{\dagger})^{(2)}_{\mu}
\label{O6p2}
\ea
is an $O(6)$ tensor with $\sigma=2$. For $h_0\neq h_2$ the Hamiltonian 
$H_2$ is neither an $O(6)$-scalar nor an $O(5)$-scalar hence can induce 
both $O(6)$ and $O(5)$ mixing subject to $\Delta\sigma=0,\pm 2$ and 
$\Delta\tau=\pm 1,\pm 3$. Although $H_2$ is not invariant under $O(6)$, 
it still has an exactly solvable ground band with good $O(6)$ symmetry. 
This arises from the fact that the boson pairs of Eqs.~(\ref{O6p0}) and 
(\ref{O6p2}) annihilate the state $\vert c;\,N\rangle$ of 
Eq.~(\ref{condeno6}), which is the $O(6)$ intrinsic state for the 
ground band with $\sigma=N$. 
Since $H_2$ is rotational invariant, states of good angular momentum $L$
projected from $\vert c;\,N\rangle$ are also eigenstates of
$H_2$ with good $O(6)$ symmetry and form its ground band. 
These projected states do not have good $O(5)$ symmetry and their
known wave functions contain a mixture of components with different $\tau$.
It follows that $H_2$
has a subset of solvable states with good $O(6)$ symmetry ($\sigma=N$),
which is not preserved by other states.
All eigenstates of $H_2$ 
break the $O(5)$ symmetry but preserve the $O(3)$ symmetry. 
These are precisely the required features of type III $O(6)$ PDS.

\begin{table}[t]
\caption{Calculated~\protect\cite{levisa02} and observed
$B(E2)$ values 
(in $10^{-2}e^2b^2)$ for $^{162}$Dy. The $E2$ operator is
$T^{(2)}= e_{B}[\,d^{\dagger}s 
+ s^{\dagger}\tilde d + \chi\, (d^{\dagger}\tilde{d}\,)^{(2)}\,]$ 
with parameters $e_B=0.138$ $[0.127]$ $eb$ and $\chi=-0.235$ $[-0.557]$ 
for $H_1$ (\ref{h1}) [$H_2$ (\ref{h2})].} 
\begin{center}
\footnotesize
\begin{tabular}{llll|llll}
\hline
Transition & $H_{1}$ & $H_{2}$ & Expt. & 
Transition & $H_{1}$ & $H_{2}$ & Expt. \\
\hline
$2^{+}_{K=0_1}\rightarrow 0^{+}_{K=0_1}$  & 107   & 107  & 107(2) &
$2^{+}_{K=2_1}\rightarrow 0^{+}_{K=0_1}$  & 2.4   & 2.4  &   2.4(1)  \\
$4^{+}_{K=0_1}\rightarrow 2^{+}_{K=0_1}$  & 151   & 152  & 151(6) &
$2^{+}_{K=2_1}\rightarrow 2^{+}_{K=0_1}$  & 3.8   & 4.0  & 4.2(2)  \\
$6^{+}_{K=0_1}\rightarrow 4^{+}_{K=0_1}$  & 163   & 165   & 157(9) &
$2^{+}_{K=2_1}\rightarrow 4^{+}_{K=0_1}$  & 0.24  & 0.26 & 0.30(2) \\
$8^{+}_{K=0_1}\rightarrow 6^{+}_{K=0_1}$  & 166   & 168 & 182(9)   &
$3^{+}_{K=2_1}\rightarrow 2^{+}_{K=0_1}$  & 4.2 & 4.3 &           \\
$10^{+}_{K=0_1}\rightarrow 8^{+}_{K=0_1}$ & 164   & 167 & 183(12)  &
$3^{+}_{K=2_1}\rightarrow 4^{+}_{K=0_1}$  & 2.2  & 2.3  &           \\
$12^{+}_{K=0_1}\rightarrow 10^{+}_{K=0_1}$& 159  & 163 & 168(21)  &
$4^{+}_{K=2_1}\rightarrow 2^{+}_{K=0_1}$ & 1.21 & 1.14 & 0.91(5)  \\
 & & & &
$4^{+}_{K=2_1}\rightarrow 4^{+}_{K=0_1}$ & 4.5  & 4.7  & 4.4(3)   \\
$0^{+}_{K=0_2}\rightarrow 2^{+}_{K=0_1}$ & 0.16  & 0.23 &         & 
$4^{+}_{K=2_1}\rightarrow 6^{+}_{K=0_1}$ & 0.59 & 0.61 & 0.63(4)  \\
$0^{+}_{K=0_2}\rightarrow 2^{+}_{K=2_1}$ & 0.14  & 17.23 &         &
$5^{+}_{K=2_1}\rightarrow 4^{+}_{K=0_1}$ & 3.4 & 3.3  & 3.3(2)   \\
$2^{+}_{K=0_2}\rightarrow 0^{+}_{K=0_1}$ & 0.02  & 0.04 &         &
$5^{+}_{K=2_1}\rightarrow 6^{+}_{K=0_1}$ & 2.9  & 3.1  & 4.0(2)   \\
$2^{+}_{K=0_2}\rightarrow 2^{+}_{K=0_1}$ & 0.04  & 0.05 &         &
$6^{+}_{K=2_1}\rightarrow 4^{+}_{K=0_1}$ & 0.84 & 0.72 & 0.63(4)  \\
$2^{+}_{K=0_2}\rightarrow 2^{+}_{K=2_1}$ & 0.03  & 3.69 &         &
$6^{+}_{K=2_1}\rightarrow 6^{+}_{K=0_1}$ & 4.5  & 4.7  & 5.0(4)   \\
 & & & & & & & \\
\hline
\end{tabular}
\end{center}
\end{table}

The spectra of $H_2$ is shown in Fig.~3, 
while the $O(6)$ and $O(5)$ decomposition of selected states are 
shown in the lower portion of Fig.~4 and Fig.~5 respectively.
For $H_2$, the solvable $K=0_1$ ground band has $\sigma=N$ 
and all eigenstates are mixed with respect to $O(5)$. 
However, in contrast to $H_1$ of Eq.~(\ref{h1}), 
excited bands of $H_2$ can have 
components with different $O(6)$ character. For example, 
the $K=0_2$ band of $H_2$ has components with $\sigma=N$ $(85.50\%)$,
$\sigma=N-2$ $(14.45\%)$, and $\sigma=N-4$ $(0.05\%)$.
These $\sigma$-admixtures can in turn be interpreted in terms of 
multi-phonon excitations. 
Specifically, we find that the $K=0_2$ band is composed of
$36.29\%$ $\beta$, $63.68\%$ $\gamma^2_{K=0}$,
and $0.03\%$ $\beta^2$ modes,
{\it i.e.}, it is dominantly a double-gamma phonon excitation
with significant single-$\beta$ phonon admixture. The $K=2_1$ band 
has only a small $O(6)$ impurity and is 
an almost pure single-gamma phonon band.
The combined results of Figs.~4 and~5 constitute a direct proof
that $H_2$ (\ref{h2}) possesses a type III $O(6)$ PDS
which is distinct from the type II $O(6)$ PDS of $H_1$ (\ref{h1}).

In Table~2 we compare the presently known experimental $B(E2)$ values
for transitions in $^{162}$Dy
with PDS calculations.
The $B(E2)$ values predicted by $H_1$ and $H_2$
for $K=0_1\rightarrow K=0_1$ and $K=2_1\rightarrow K=0_1$ transitions
are very similar and agree well with the measured values.
On the other hand, their predictions for interband transitions
from the $K=0_2$ band are very different. Future measurements of these 
transitions will enable one to distinguish which type of partial 
$O(6)$ symmetry is more suitable for $^{162}$Dy.

\section{Summary and Conclusions}

In this contribution we have considered departures from 
complete dynamical symmetry by introducing the notion of a partial 
dynamical symmetry (PDS). The latter refers to an intermediate symmetry 
structure for which some (but not all) of the virtues 
of a dynamical symmetry ({\it e.g.} solvability, quantum numbers) 
are retained. We have presented empirical examples of nuclei 
in each category of PDS. Although we have focused the discussion to 
partial symmetries in systems 
of one type of bosons (IBM-1) relevant to nuclei, 
there are also examples of PDS in systems of several types of bosons 
\cite{talmi97,levgin00} ({\it e.g.} proton-neutron bosons in the IBM-2), 
in bose-fermi systems \cite{jolos00} (IBFM) and in purely fermionic 
systems~\cite{esclev00,rowe01}. Thus, PDS seem to be a generic feature in 
dynamical systems with concrete applications to nuclear and 
molecular \cite{pinchen97} spectroscopy. In addition, PDS have been shown 
to be relevant to the study of mixed systems~\cite{whelan93,levwhe96} 
with coexisting regularity and chaos.

\section*{Acknowledgments}

It is a pleasure and honor for me to dedicate this contribution to F. 
Iachello on the occasion of his 60th birthday. 
Franco's innovative approach to physics, emphasizing the unity of science 
and uncovering of underlying symmetries, 
has influenced and inspired the ideas discussed here. 
Segments of the reported results were obtained in collaboration with 
I. Sinai (HU), P. Van Isacker (GANIL) and 
J.E. Garcia-Ramos (Huelva). 
This work was supported by the Israel Science Foundation.

\end{document}